\documentclass[aps,prl,twocolumn,groupedaddress,showpacs]{revtex4}
\usepackage{float}
\usepackage{dcolumn}
\usepackage{amsmath}
\usepackage{amssymb}
\input{epsf}

\ifx\pdftexversion\undefined
  \usepackage{graphicx}
\else
  \usepackage[pdftex]{graphicx}
\fi

\newtheorem{theorem}{Theorem}

\newtheorem{claim}[theorem]{Claim}

\newcommand{\qedsymb}{\hfill{\rule{2mm}{2mm}}}
\newenvironment{proof}[1][]{\begin{trivlist}
\item[\hspace{\labelsep}{\bf\noindent Proof#1:\/}] }{\qedsymb\end{trivlist}}

\newcommand{\ket}[1]{\left| {#1} \right\rangle}
\newcommand{\bra}[1]{\left\langle {#1}\right |}

\newcommand\abs[1]{{\left| {#1} \right|}}

\def\p{\ket{\psi}}

\def\be{\begin{equation}}
\def\ee{\end{equation}}
\def\bea{\begin{eqnarray}}
\def\eea{\end{eqnarray}}

\def\>{\rangle}
\def\<{\langle}

\begin{document}

\title{From quantum circuits to adiabatic algorithms}

\author{M. Stewart Siu\footnote{msiu@stanford.edu}}

\affiliation{Department of Physics, Stanford University, Stanford CA 94305}

\date{\today}

\begin{abstract}
This paper explores several aspects of the adiabatic quantum computation model. We first show 
a way that directly maps any arbitrary circuit in the standard quantum computing model to an 
adiabatic algorithm of the same depth. Specifically, we look for a smooth time-dependent 
Hamiltonian whose unique ground state slowly changes from the initial state of the circuit to 
its final state. Since this construction requires in general an n-local Hamiltonian, we will 
study whether approximation is possible using previous results on ground state entanglement 
and perturbation theory. Finally we will point out how the adiabatic model can be relaxed in 
various ways to allow for 2-local partially adiabatic algorithms as well as 2-local holonomic 
quantum algorithms.

\end{abstract}

\maketitle


\section{1. Introduction}

Adiabatic evolution as a quantum computation model has attracted much attention since its 
introduction by Farhi et al \cite{fggs}. The basic idea is the following: Start with a 
Hamiltonian whose ground state is easily reachable and prepare our state in the ground state. 
Change it slowly to a new Hamiltonian that encodes the solution of the problem and maintain a 
large energy gap between the ground state and the excited state that the evolving state 
couples to. The Adiabatic Theorem \cite{messiah} then guarantees that the resulted state will 
be very close to the ground state of the new Hamiltonian. The original form of the Hamiltonian 
considered in \cite{fggs} is a straight-line interpolation: 
$H(s)=(1-s)H_{initial}+sH_{final}$. Recently, it was proved that any standard quantum circuit, 
specified by a sequence of unitary operators, can be implemented as an adiabatic evolution of 
this form\cite{adkll,kkr}. The authors use computational complexity techniques developed for 
proving the QMA-completeness of the k-local Hamiltonian problem (Kempe et al \cite{kkr} 
achieved the case for k=2); the evolving state encodes the entire computational history. 
Roughly speaking, they construct a Hamiltonian whose ground state is the superposition of all 
the stages in a given circuit. If the circuit has depth L, the time required to obtain this 
ground state is $O(1/L^6)$ for the 3-local Hamiltonian and there is a $O(1/L)$ probability of 
obtaining the final state of the circuit given this superposition.
On a seemingly unrelated note, Farhi et al \cite{fgg} showed after \cite{fggs} that if we do 
not restrict adiabatic evolution to the straight-line path and add terms that vanish at the 
endpoints, we may be able to turn an inefficient computation into an efficient one. A general 
method for finding an efficient path is however not known. In light of these two developments, 
we may ask - 
Can we always find an efficient adiabatic evolution path, not necessarily of the straight-line 
fom, for problems efficiently solvable by quantum circuits such that we directly obtain the 
desired final state? Starting with this question, we will present several variations on how to 
implement the standard circuit model by adiabatic evolution.  

The first result we show is that once we specify a) the unitary transformation that takes the 
eigenstates at the beginning to those at the end of the evolution, and b) how we want the 
eigenvalues to evolve, we can immediately derive the form of time-dependent Hamiltonian 
required without the use of any ancilla qubits. Conceptually the simplest example is a 
time-dependent similarity transform, $U(t)H_{initial}U(t)^\dag$. This observation should allow 
us to engineer Hamiltonians according to computational needs. However, a Hamiltonian of the 
type $...U_2U_1H_{initial}U_1^\dag U_2^\dag...$ can be highly non-local even if $U_1,U_2...$ 
and $H_{initial}$ have simple local forms. Since we are interested in the ground state, we 
should ask whether it is possible to find an local approximation. 

It turns out that while approximations are possible to a certain extent, there is much 
constraint. We will demonstrate this point in two steps. First we will make use of the results 
by Haselgrove et al\cite{hno}, which show how the entanglement of the eigenstates of a 
Hamiltonian is related to what bodies in the system each term in the Hamiltonian acts 
non-trivially on. Intuitively speaking, if an eigenstate shows strong correlation between 
bodies which the Hamiltonian does not directly couple, i.e. act nontrivially on all as a 
tensor product, the Hamiltonian cannot distinguish very well between such a state and other 
similarly entangled states that are orthogonal to it. This results in a small energy gap. 
Since a quantum circuit can generate highly correlated states, when we want to make them 
ground states of a Hamiltonian in an adiabatic algorithm, they will be difficult to 
approximate. Then, as an explicit example, we will use the approximation method developed in 
\cite{kkr}, derived from perturbation theory, and apply it on our construction. We will see 
that we could indeed make a local approximation under the constraint implied by \cite{hno}, 
but the resulting evolution can be inefficient.

Next we look at how this approach of transforming the Hamiltonian adiabatically is related to 
the manipulation of geometric phase. We start by asking, given local approximation is 
difficult: Why is the adiabatic model more demanding than the basic circuit model, for which 
2-local Hamiltonians easily suffice with $U=exp(iH_{2-local}t)$ for each gate? There are at 
least two crucial differences between the two models. 

\begin{itemize}
\item[I)] The adiabatic model keeps track of exactly where the state is at every moment 
throughout the evolution and penalizes any deviation, while the circuit model keeps no 
information about the state at all. This makes the former more resistant to error. \\ 
\item[II)] The adiabatic model allows time variability in the application of the Hamiltonian, 
while the basic circuit model requires precise pulse timing. 
\end{itemize}
The word "adiabatic" itself only suggests property II) above, so if we are willing to relax 
property I), we would have much more freedom to design our Hamiltonian. The main issue we need 
to deal with, as we will show, is the geometric phase. Suppose we implement the Hamiltonian 
$U(s)H_{initial}U(s)^\dag$ without making sure that we start with the ground state or any 
eigenstate of the initial Hamiltonian. Instead of having $U(s)$ applied on our initial state, 
there will be further transformation due to the relative phases accumulated between different 
eigenstates. The dynamical component of the phase is straightforward to cancel out, but the 
geometric component is more subtle to calculate. To avoid having to cancel the geometric 
component, we may either make sure we always start with an eigenstate of the Hamiltonian, or 
take advantage of the geometric phase to implement the desired transformation. This would 
naturally lead us to the holonomic quantum computing model developed by Zanardi et 
al\cite{holo1}, which in fact precedes the adiabatic algorithm of \cite{fgg}. 

Let us give a lightning review of the idea of holonomic quantum computing (HQC). Wilzcek and 
Zee introduced in \cite{wz} the observation that if a Hamiltonian with degenerate eigenstates 
goes through a cycle adiabatically without changing the degeneracy of each level, the 
degenerate subspace can be viewed as a gauge group on the manifold corresponding to the 
parameter space of the Hamiltonian. After each cyclic evolution, an arbitrary state in the 
degenerate space will undergo an unitary transformation depending on the path taken; the set 
of all possible such unitary transformation given a parameter space that specifies the 
Hamiltonian is called the holonomy group, and the parameter space is often called the control 
manifold. Elements of the group generally do not commute, so the transformation is called the 
non-Abelian geometric phase. Zanardi et al \cite{holo1} applied this idea on quantum computing 
by choosing initial Hamiltonians for which the computational states are completely degenerate. 
Transformations are then applied by holonomy. In addition to time variability, the geometric 
nature (such as dependency on the area of the loop) also gives HQC some resistance to errors.   

Seeing HQC as a generalized adiabatic model brings us many new insights. First, we can apply 
the local approximation techniques of \cite{kkr} to show that 2-local Hamiltonians are 
sufficient to implement HQC. Furthermore, we show that the construction of \cite{adkll} etc., 
originally developed for computational complexity proofs, has a hidden gauge freedom and can 
be viewed as half a holonomic cycle. This view allows us to improve the adiabatic 
implementation so that we obtain only the desired final state instead of the computational 
history.

This paper is organized as follows. In section 2, we give the direct way to construct an 
adiabatic equivalent of any circuit without encoding the computational history. It makes use 
of only the same number of qubits as in the circuit and a running time of the same order as 
the depth of the circuit, as shown in section 3. This in general involves n-local Hamiltonian, 
and section 4 discusses why this can be difficult to approximate by studying entanglement 
properties of the ground state of Hamiltonians. In section 5 we show one way to construct a 
2-local Hamiltonian whose worst case run-time scales exponentially with n, illustrating a 
tradeoff between resource requirement and running time. In section 6, we look at how 2-local 
constructions can be useful for generalized adiabatic algorithms and how computation models 
that use Abelian and non-Abelian geometric phases \cite{geom, holo1} fall into this category. 
Finally, we return in section 7 to the computational history approach of \cite{adkll} and note 
its interesting connection to holonomic quantum computing.

\section{2. A direct mapping}

We adopt a general definition of adiabatic computation and look for a time-dependent, 
differentiable Hamiltonian $H(s)$, where $0<s<1$ is the time parameter, such that $H(0)$ is an 
initial Hamiltonian with a unique, easily reachable ground state and $H(1)$ is a Hamiltonian 
with a unique ground state encoding the solution of our problem.  A quantum circuit can be 
given in the form $\p=U_lU_{l-1}...U_1\ket{0}$, where $U_i$ are unitary operators representing 
one or two qubit gates. To map this transformation into adiabatic evolution, we start with a 
Hamiltonian $H(0)$, whose ground state is $\ket{0}$, and we would like to have $H(s)$ such 
that $\p$ is the ground state of $H(1)$. The most common problem in constructing such an 
$H(s)$ is that the energy gap between the ground state and the first excited state varies 
during the evolution. A small gap implies a larger probability for the ground state to be 
excited, and in turn a longer evolution time if we want to compensate for it.

Our main observation is that it is possible to maintain a constant gap size as long as we keep 
the Hamiltonian $H(s)$ to be of the form $U(s)H(0)U^\dag(s)$. Let us be more specific. Suppose 
the circuit requires us to perform unitary gate $U$ on state $\ket{0}$. Let $K=-ilogU$ and 
$\tilde{U}(s)=exp(isK)$, such that $\tilde{U}(0)=1$ and $\tilde{U}(1)=U$. We start with a 
Hamiltonian $H$ with $\ket{0}$ as its ground state:
\be
H\ket{n}=E_n\ket{n}
\ee
We can add $V(s)$ such that the following is true:
\be
(H+V(s))\tilde{U}(s)\ket{n}=f(n,s)\tilde{U}(s)\ket{n}
\ee
if
\be
V(s)\tilde{U}(s)\ket{n}=[\tilde{U}(s),H]\ket{n} + (f(n,s)-E_n)\tilde{U}(s)\ket{n}
\label{commutator}
\ee
This completely specifies $V(s)$, and if $f(n,s)=E_n$, $V(s)$ is just, in the original 
(computational) basis, $\tilde{U}(s)H\tilde{U}(s)^\dag - H$. It is clear that as $s$ goes to 1 
slowly, we obtain $U\ket{0}$ as our ground state without worrying about a shrinking gap. Note 
that $f(n,s)$ allows us to manipulate the gap size. 

Using the idea above, we can now spell out the explicit mapping. Given $U_1,...U_l$, we first 
replace the overall time parameter $s$ by a series of time step parameters $s_i$ for $i=1..l$, 
$s_i\subset[0,1]$. This means:
\be
H(s)=(\prod_{i=l..1}\tilde{U}(s_i))H(0)(\prod_{i=1..l}\tilde{U}^\dag(s_i))
\label{bigham}
\ee

Let the Hamiltonian at the beginning of the i-th time step be $H^{(i-1)}=\sum_j 
h^{(i)}_j=\sum_j^{\parallel} h^{(i)}_j+\sum_j^{\perp} h^{(i)}_j$ where $h^{(i)}_j$ denotes 
individual local Hamiltonians. $\sum^{\parallel}$ and $\sum^{\perp}$ refer respectively to 
terms whose qubits overlap with those of $U_{i}$ and terms that act on different qubits. In 
this notation, we can write $V(s_i)$ as
\be
V(s_i)=\tilde{U}_i(s_i) ( \Sigma_j^{\parallel} h^{(i)}_j ) \tilde{U}^\dag_i (s_i) - 
\Sigma_j^{\parallel} h^{(i)}_j   
\label{vform1}
\ee
For illustrative purpose, let us consider a typical term, where $U_i$ is the controlled-Z gate 
(which with single-qubit gates is universal) acting on the first two qubits, and $h^{(i)}_j$ 
acts on the second qubit as well as some other qubits. The matrix representation of 
$h^{(i)}_j$ and $U_i$ for the first two qubits looks like
\[
h^{(i)}_j=\left(\begin{array}{cccc}
h_1 & h_2 &     &     \\
h_3 & h_4 &     &     \\
    &     & h_1 & h_2 \\ 
    &     & h_3 & h_4 \\
	\end{array}\right),
U_i = \left(\begin{array}{cccc}
1& & & \\
 &1& & \\
 & &1& \\
 & & &-1\\
 \end{array}\right)	
\]

Then $\tilde{U}_i(s_i) h^{(i)}_j \tilde{U}_i^\dag(s_i) - h^{(i)}_j =$
\begin{eqnarray}
\left(\begin{array}{cccc}
0& & & \\
 &0& & \\
 & &0&(e^{-is\pi-1})h_2 \\
 & &(e^{is\pi-1})h_3& 0\\
 \end{array}\right)	
 \label{cz}
\end{eqnarray}
Note from this example that if $H^{(i-1)}=\sum h^{(i)}_j$ is $m$-local, $V(s_i)$ can be at 
most $(m+1)$-local, and this happens when exactly one qubit of a two qubit gate $U_i$ overlaps 
with one qubit of $h^{(i)}_j$. Thus $V(s_i)$ can be up to $n$-local where $n$ is the total 
number of qubits. We will study more closely the complexity and locality of such Hamiltonians 
in section 4.      

Let us look at another specific example using Pauli matrices X, Y and Z as basis.
Suppose for two qubits we start the Hamiltonian:
\be
H = ZZ - ZI + IZ
\ee
where $ZI$ means a $Z$ on the first qubit and identity on the second qubit etc. Clearly the
ground state is $\ket{10}$. With that as our starting point, we can apply a CNOT and
see how it turns into $\ket{11}$. The recipe above tells us the Hamiltonian we need to add is
\begin{eqnarray}
V(s) = & sin(s\pi) IY + (1-cos(s\pi)) IZ \nonumber \\ 
& - sin(s\pi) YZ - (1-cos(s\pi)) ZZ
\end{eqnarray}
We can see that as s goes to 1, the new Hamiltonian will become
$H+V(1)$=$-ZZ+ZI+IZ$, whose ground state is indeed $\ket{11}$. The $IY$ and $YZ$ terms are 
zero at the end points, as the extra terms in \cite{fgg} are. 

\section{3. Error bounds}

We now check the evolution time required for each step according to the Adiabatic Theorem 
\cite{messiah}. The result here is useful for the construction in section 5 and 6 as well. 
Under the adiabatic approximation, the evolving state is proportional to the instantaneous 
eigenstate of the time-dependent Hamiltonian. Substituting this into Schrodinger's equation, 
this means the time derivative does not take one eigenstate to another, i.e. 
$\bra{m,s_i}{d\over dt_i}\ket{0,s_i}\sim 0$, $m\neq 0$. Thus the correction to the 
approximation must be proportional to (we define $K_i=-i log U_i$ below)
\begin{eqnarray}
\alpha(s_i)&\sim&\sum_{m\neq 0}\bra{m,s_i}{d\over dt_i}\ket{0,s_i} \nonumber \\
             &=&{1\over T}\sum_{m\neq 
0}\bra{m,s_i}{d(\tilde{U}(s_i)H^{(i-1)}\tilde{U}^\dag(s_i))\over (E_m-E_0)ds_i}\ket{0,s_i} 
\nonumber \\
&=&{1\over T}\sum_{m\neq 0}\bra{m,s_i}e^{is_iK_i}{[K_i,H^{(i-1)}]\over 
E_m-E_0}e^{-is_iK_i}\ket{0,s_i} \nonumber \\
&=&{1\over T}\sum_{m\neq 0}\bra{m,s_i=0}{[K_i,H^{(i-1)}]\over E_m-E_0}\ket{0,s_i=0} \nonumber 
\\
&=&{1\over T}\sum_{m\neq 0}-\bra{m,s_i=0}K_i\ket{0,s_i=0} \label{error} 
\end{eqnarray} 
where $\ket{m,s_i}$ denotes the instantaneous eigenstate with eigenvalue $E_m$ and 
$s_i=t_i/T$. $H^{(i-1)}$ preserves the spectrum of $\ket{m,s_i=0}$, so the contribution to the 
above term is due to $K_i$. Taking $U_i$ to be controlled-Z as an example again, the 
eigenvalues of $K_i$ are 0 and $\pi$. $\alpha(s_i)$ is therefore bounded by $\pi/T$. The total 
time required for the step is proportional to the transition probability to other states, 
which according to \cite{messiah}, is bounded by $\abs{\hbar \alpha(s_i)\over E_m-E_0}^2$ for 
the smallest $E_m$. Remarkably, the error is not only independent of total number of qubits 
$n$, it is also independent of $s_i$, which means further local variation in evolution speed 
is not required to achieve optimal timing. Of course, $\tilde{U}(s_i)=exp(isK)$ is just one 
arbitrary choice we make; there may be other forms of $\tilde{U}(s_i)$ that yield better 
performance or are easier to implement. We should note that it is possible to eliminate the 
error altogether by adding auxillary terms to the Hamiltonian, but this would only be useful 
for state preparation as it generally requires complete knowledge of what we want to generate.

\section{4. Locality of the Hamiltonian}

In hindsight it should not be surprising that this direct mapping yields an n-local 
Hamiltonian. After all, while it is easy to decompose an n-local unitary operator into a 
product of 2-local ones, since 2-qubit gates are universal, it is far more difficult to 
approximate an n-local operator with a sum of 2-local operators, even with the addition of 
ancilla qubits. This section is devoted to the understanding of this difficulty. 

First we review some results by Haselgrove et al \cite{hno}. In \cite{hno} the authors show 
how the entanglement of the eigenstates of a Hamiltonian is related to its coupling topology. 
Intuitively speaking, if an eigenstate shows strong correlation between bodies which the 
Hamiltonian does not directly couple (i.e. act nontrivially on all of them as a tensor 
product), the Hamiltonian cannot distinguish very well between such a state and other 
similarly entangled states that are orthogonal to it. This results in a small energy gap. The 
following theorem from \cite{hno} makes this idea concrete and suffices for our purpose.

\begin{theorem}
Consider a state $\p$ and a Hamiltonian H whose eigenvalues and eigenstates are $E_j$ and 
$\ket{E_j}$ respectively with j=0..d-1; d is the dimensional of the Hilbert space and 
$\ket{E_0}$ is the ground state of H. Let F be the overlap of $\p$ with $\ket{E_0}$ and 
$E_{tot}$ be the difference between the maximum and minimum eigenvalues. Then for all density 
matrices $\rho$ with eigenvalues $\rho_1 \leq \rho_2 \leq \rho_3...$, such that $tr(\rho 
H)$=$\bra{\psi}H\p$, the following inequality holds: 
\be
\sum_{j=1}^{d-1}(E_j-E_0)\rho_{j+1}\leq (1-F^2)E_{tot}
\label{ineq}
\ee
\end{theorem}

The proof is elementary and we will refer the readers to the lucid explanation in \cite{hno}. 
Now we may apply this theorem on the construction in section 2.

\begin{claim}
Let $H_0$ be a 1-local Hamiltonian with unique ground state $\ket{0}$. There does not exist 
general $k$-local approximation for the n-local Hamiltonian $UH_0U^\dag$, where $k \leq n-1$ 
and U is a polynomial-sized circuit, such that the approximation produces exactly the same 
ground state and first excited state. Specifically, one cannot always construct a $k$-local 
Hamiltonian which has $U\ket{0}$ as a non-degenerate eigenstate. 
\end{claim}

\begin{proof}

We will start with the case without ancilla qubits. Consider the state $\p$=${1 \over 
\sqrt{2}}(\ket{000...}+\ket{111...})$,the n-qubit GHZ state. Consider also a k-local 
Hamiltonian H whose ground state $\ket{E_0}$=$\p$, so F=1. If we choose 
\be
\rho = {1 \over 2} (\ket{000...}\bra{000...} + \ket{111...}\bra{111...})
\ee
where "..." again indicates n zeroes or ones, it is easy to see that 
$tr_{n-k}\rho$=$tr_{n-k}\p\bra{\psi}$ for $k\leq n-1$, where $tr_{n-k}$ means tracing over any 
n-k qubits. It then follows that for $k\leq n-1$, $tr(\rho H)$=$\bra{\psi}H\p$ for a k-local 
Hamiltonian H. Putting this into the inequality in Theorem 1, we obtain $E_1-E_0$=0, meaning 
that the ground state corresponding to $E_0$ is degenerate. 

Now suppose a k-local exact approximation exists. Choose the n-qubit polynomial sized circuit, 
U=$Hadamard_1 \prod_{i=1}^{i=n-1} CNOT_{i,i+1}$, i.e. a Hadamard gate acts on the first qubit, 
followed by a series of CNOTs on the first and second, the second and the third, and so on. 
Clearly, this circuit acting on the initial state $\ket{000...}$ produces $\p$. Suppose we 
start with a simple 1-local Hamiltonian $H_0$ (e.g. Set $H_0$ = $\sum_{i=1}^n \sigma_{z}^{i}$) 
which has $\ket{000...}$ as a non-degenerate ground state. If there exists a procedure that 
exactly approximate $U H_0 U^\dag$ with a k-local Hamiltonian, $k \leq n-1$, this implies 
there exists a k-local Hamiltonian which has $\p$ as a non-degenerate ground state. Hence we 
have arrived at a contradiction.

To generalize this to the case with ancilla qubits, a slight extension of Theorem 1 is needed. 
Let the ground states of the k-local approximation be $\p\ket{a_j}$, where $\ket{a_j}$, j = 
1..m enumerates the degeneracies due to the ancilla qubits. This product form is necessary if 
we want the computational qubits to remain as $\p$. Let $\rho^{'}=\sum_{j=1}^m 
\rho\otimes\ket{a_j} \bra{a_j} /m$. Setting $E_0$ = 0, it is easy to check that $tr(\rho^{'} 
H) = \sum_{j=1}^m tr ( \rho\otimes\ket{a_j} \bra{a_j} /m) = 0$, which would force the ground 
state degeneracy to be 2m. This in turn implies that there must be degenerate ground states 
due to states orthogonal to $\p$, contradicting the assumption that there is a k-local 
Hamiltonian with $\p\ket{a_j}$ as the only ground states. 

\end{proof}

So far we have seen that a local Hamiltonian cannot have certain states as its ground state, 
as shown by \cite{hno, hno2}. This is rather expected as it is well-known \cite{reducedstates} 
that there are quantum states not determined by any reduced density matrices. The more 
interesting connection we would like to point out here, however, is the tradeoff between 
proximity to a non-local state and the energy gap, as apparent in Theorem 1. Since the energy 
gap condition is essential to adiabatic algorithms (while some forms of adiabatic theorem 
without gap condition exists, they cannot guarantee the final state to arbitrary accuracy 
\cite{gap}), this places another direct tradeoff between accuracy and running-time. Now that 
an exact approximation is not possible, we will look at how close we can get.

In \cite{kkr}, a 2-local approximation for 3-local Hamiltonians is contructed (see section 5). 
Normalizing the total energy to unity, the ground state energy gap for the 2-local Hamiltonian 
scales as $\delta^3$ for a ground state $O(\delta)$ close to the original ground state. In 
fact we can use Theorem 1 to make this more precise: If the energy gap scales as $\delta^3$, 
the ground state for the 2-local Hamiltonian has to be at least $O(\delta^3)$ away from an 
original GHZ-type ground state. This proves that there does not exist an approximation scheme 
better than \cite{kkr} in such a way that the energy gap scales, say, logarithmically (i.e. 
$O(1/log \delta^{-1}))$ instead of polynomially with the accuracy $O(\delta)$ of the ground 
state.

Following this idea, we can place some bounds on how good the approximation for an n-local 
Hamiltonian can be. For simplicity we will consider a 2-local approximation, should one exist, 
that has a unique ground state, ancilla qubits included. Consider the state $\ket{\phi}$:
\be
\ket{\phi}={1\over 2}(\ket{000}+\ket{111})\otimes{1\over 2}(\ket{000}+\ket{111})\otimes...
\ee
which is a tensor product of mostly 3-qubit GHZ states. It is not difficult to see that there 
are $\approx 2^{n/3}$ orthogonal states to it there are not distinguishable by 2-local terms. 
Thus we can form a density operator of rank $\approx 2^{n/3}$ and substitute it into the 
inequality (\ref{ineq}). This tells us that the average energy of these $\approx 2^{n/3}$ 
states has a gap with the ground state that is at most $(1-F^2)\sim O(\delta)$. We can tighten 
this bound a little by considering the distribution of states. If we start with the 1-local 
Hamiltonian $H_0$ = $\sum_{i=1}^n \sigma_{z}^{i}$ (the minimal form required for a unique 
ground state), $UH_0 U^\dag$ has n eigenvalues $E_j=j/n$ with degeneracy $n!/[j!(n-j)!]$. 
Simple counting shows that for the lowest $\approx 2^{n/3}$ states, the average energy is at 
least $\approx nE_1/6$. Thus we can tighten the bound to $E_1-E_0 < O(\delta/n)$.

Before we conclude this section, we should briefly note another line of attack due to 
\cite{hno2}. Generalizing beyond the GHZ-type states, states corresponding to non-degenerate 
quantum error correction codes (QECC) also turn out to be interesting for the study of local 
Hamiltonians. These are states with the property that, for some constant $t$ usually much 
smaller than n, any Pauli matrx operators acting non-trivially on up to t-qubits will take the 
state to a set of orthogonal states. Therefore, for a QECC state $\ket{x}$, if the operator 
$H-EI$ is t-local, $(H-EI)\ket{x}$ will be a sum of orthogonal states, implying that $\| 
(H-EI)\ket{x} \|$ cannot be close to zero. This prevents any QECC states from being even close 
to any eigenstate of a t-local Hamiltonian. Recasting this result in our language, we see that 
there does not exist a k-local approximation to arbitrary n-local Hamiltonians $UH_0U^\dag$ 
for sufficiently large n \em without ancilla qubits \em, because QECC states can also be 
generated efficiently by quantum circuits (see references in \cite{hno2}). With ancilla 
qubits, however, cancellation can occur for $(H-EI)\ket{x}\ket{a}$ if $\ket{a}$ is not a QECC 
state. We obtain instead a set of constraint equations that the approximation Hamiltonian has 
to satisfy in order to produce QECC states as an eigenstate.   

We hope that the discussion above would be useful for further research on not only the 
possibility of local approximation, but also the connection between local properties of 
Hamiltonians and polynomial-sized quantum circuits.

\section{5. A local approximation using the three-qubit gadget}

After an abstract discussion of possible local approximations, we will now look at explicitly 
how an approximation scheme can be used in an adiabatic algorithm. We will use as an example 
the 3 to 2-local reduction introduced by \cite{kkr}, referred to as the three-qubit gadget 
from now on. 

Let us begin with some general considerations. In order to directly map a quantum circuit to 
an adiabatic algorithm gate by gate with some approximation every time, we would need L 
approximations where L is the depth of the circuit. To achieve an error within $O(\epsilon)$ 
for the final state, including the error due to adiabatic approximation discussed in section 
3, we can estimate the required accuracy at each step as the following. Consider the worst 
case scenario, when all the errors accumulated are in the same direction. We first express the 
angle between the correct final state and the approximate final state as $\theta = 
O(\sqrt{\epsilon})$ for small $\theta$. The average angle accumulated at each step is 
$\theta/L$ because the unitary gates preserve angles. The allowed error at each step is 
therefore $1-cos^2(\theta/L)=O(\epsilon/L^2)$. Hence as long as the energy gap size scales 
polynomially with this allowed error, the adiabatic algorithm is efficient. We will see one 
such example in section 6.   

At every step, however, if we repeatedly apply the same approximation procedure on the 
approximate Hamiltonian from the previous step, the energy gap size would generally not scale 
polynomially with the allowed error. This is because, as observed in section 4, the energy gap 
will have to be scaled down by at least a factor of $O(\delta)$ for an allowed error of 
$O(\delta)$. Repeatedly approximating approximate Hamiltonians thus results in an energy gap 
of at most $O(\delta^L)$. This would hold true for any schemes. 

We may now look at the specific scheme based on \cite{kkr}. The authors develop a framework of 
perturbation theory that gives sufficient conditions for how one Hamiltonian can approximate 
another. The basic idea they consider is as follows. A 3-local Hamiltonian $H_{3}$ can be 
represented as a 2-local Hamiltonian restricted to a certain subspace, the intuition being 
that when the interaction involves more bodies, we have finer restrictions on the eigenspaces. 
Let this 2-local Hamiltonian be $V_2$ and the subspace be $S$. If we add another 2-local 
Hamiltonian $H_2$, such that $H_2$ is zero on S and large everywhere else, it is not difficult 
to see that the lower spectrum of $\tilde{H}_2=H_2+V_2$ is close to that of $H_3$, as $H_2$ 
gives penalty to states outside of S and restricts $V_2$ to $S$.

With this intuition, the next tool we need is a good measure of the lower spectrum of 
$\tilde{H}_2$. This is provided by the self-energy $\Sigma_- (z)$ (analogous to the sum of one 
particle irreducible diagrams in field theory) defined as follows. First we define the Green 
function $\tilde{G}(z)$ of $\tilde{H}_2$ as 
\be
\tilde{G}(z)=(zI-\tilde{H}_2)^{-1}
\ee
Now we define $\Sigma_{-}(z)$ by
\be
\tilde{G}_{--}(z)=(z I_{-}-\Sigma_{-}(z))^{-1} 
\ee
where $\tilde{G}_{--}(z)$ is $\tilde{G}(z)$ restricted to the lower spectrum of $H_2$ (not 
$\tilde{H_2}$!). With this definition, \cite{kkr} proved that (Theorem 4, Lemma 9) if 
\begin{equation}
\|\Sigma_{-}(z)-H_{eff}\|\le\delta 
\label{sigma}
\end{equation}
for some operator $H_{eff}$, then both the lower eigenvalues and the ground states of 
$\tilde{H}_2$ will be $\cal{O}$$(\delta)$ close to $H_{eff}$. From this result, we would have 
a good approximation for $H_3$ if there is an $H_{eff}$ that is manifestly the same as $H_3$ 
on the computational qubits in the energy range we are interested in.

For any 3-local term $H_3$, Kempe et al propose an $H_2$ on ancilla qubits and a $V_2$ 
coupling the computational qubits with ancilla qubits, such that when we calculate 
$\Sigma_{-}$, the above equation is satisfied. This construction is called a three-qubit 
gadget. To apply this to our adiabatic algorithm, we note from section 2 that for each 2-qubit 
quantum gate we add to the Hamiltonian, a m-local Hamiltonian can become at most m+1-local. 
This means if we start with an 1 or 2-local Hamiltonian and apply the three-qubit gadget at 
every step, we should arrive at a 2-local Hamiltonian at the end. 
Let us write out the terms explicitly:

To begin with, the following Hamiltonian on the ancilla qubits (playing the role of $H_2$ 
above) is added:
\be
H_{anc}=-\frac{\delta^{-3}}{4} \sum_{i=1}^l \sum_m I\otimes 
\bigl(\sigma^{z}_{im1}\sigma^{z}_{im2} +\sigma^{z}_{im1}\sigma^{z}_{im3}
+\sigma^{z}_{im2}\sigma^{z}_{im3}-3I\bigr) 
\ee

Terms like $\sigma^{z}_{im1}$ are Pauli matrices on ancilla qubits identified by three 
indices: $i$ corresponds to the time step which runs from 1 to $l$; the meaning of the second 
and third indices will become clear shortly. $\delta$ would become the error of the 2-local 
approximation; a smaller $\delta$ would correspond to better approximated spectrum and ground 
state. Next we give an stepwise approximation, such that given a 2-local Hamiltonian 
$H^{(i-1)}$ at the beginning of each time step (see section 2), we find a 2-local perturbation 
$V'(s_i)$ to approximate the possibly 3-local $V(s_i)$ (=$U_iH^{(i-1)}U_i^\dag-H^{(i-1)}$) 
when $U_i$ is applied. To do this, we first write (\ref{vform1}) in the following form:
\begin{eqnarray}      
V(s_i) &=& \tilde{U}_i(s_i) ( \Sigma_i^{\parallel} h^{(i)}_j ) \tilde{U}^\dag_i (s_i) - 
\Sigma_i^{\parallel} h^{(i)}_j \nonumber \\
       &=& Y_i - 6 \sum_m B_{im1}B_{im2}B_{im3} \label{vform2} 
\end{eqnarray}
where $Y_i$ is 2-local and the $B$'s are positive semidefinite commuting operator acting on 
three different qubits. This decomposition is always possible because the Pauli matrix product 
$\sigma^\alpha \otimes \sigma^\beta \otimes \sigma^\gamma$ forms a basis for 3-local matrices. 
If the coefficient of a term is positive, we can rewrite the basis term as $ 
(1+\sigma^{\alpha})\otimes(1+\sigma^{\beta})\otimes(1+\sigma^{\gamma})$ + 2-local terms; if it 
is negative, we can use rewrite it as 
$-(1-\sigma^{\alpha})\otimes(1+\sigma^{\beta})\otimes(1+\sigma^{\gamma})$ + 2-local terms. 
This way we arrive at the form of \eqref{vform2}, and we can see that $m$ is the number of 
such product terms in the decomposition. Note that while this decomposition may not be obvious 
in practice, it is a constructive procedure that can be done with a classical computer 
program. Now we can construct $V'(s_i)$:
\begin{eqnarray}
V'(s_i)&=&Y_i + \sum_m \bigl\{ \delta^{-1} (B_{m1}^{2}+B_{m2}^{2}+B_{m3}^{2}) \nonumber \\
& &-~\delta^{-2}(B_{im1}\otimes\sigma^{x}_{im1} + B_{im2}\otimes\sigma^{x}_{im2} \nonumber \\
& &+~B_{im3}\otimes\sigma^{x}_{im3}) \bigr\} 
\label{vprime}
\end{eqnarray}
where the Pauli matrices in the last sum act on the ancilla qubits. Each term in the sum 
involving three ancillae is a three-qubit gadget. In summary, our total Hamiltonian is 
$H_{anc}+H(0)+\sum_i V'(s_i)$, and the error introduced in this 2-local approximation at each 
time step is $\cal{O}$$(\delta)$.

To check that $\Sigma_{-}$ satisfies equation (\ref{sigma}), put $H_2=H_{anc}$, $V_2 = 
V'(s_i)$, and expand $\Sigma_{-}$ as 
\begin{eqnarray*}
\Sigma_{-}(z) &=& V_{--} + (z-\Delta)^{-1} V_{-+}V_{+-} + (z-\Delta)^{-2} V_{-+}V_{++}V_{+-} 
\\ 
&+& (z-\Delta)^{-3} V_{-+}V_{++}V_{++}V_{+-} + \ldots
\end{eqnarray*}
where $\Delta$ is the gap of $H_2$ and $V_{+-}$ denotes the part of $V_2$ that couples the 
lower spectrum to the upper spectrum etc. We can obtain, after some algebra,
\begin{eqnarray}
\Sigma_{-}(z) &=& Y_i\otimes I_{anc} \, \nonumber \\ 
&-& \,6\sum_{m=1}^{M} B_{im1}B_{im2}B_{im3} \otimes 
\bigl(\sigma^{x}_{im1}\otimes\sigma^{x}_{im2}\otimes\sigma^{x}_{im3}\bigr) \nonumber \\ 
&+& \cal{O}(\delta).
\end{eqnarray}

Since the $B$'s are semi-positive definite, the lowest eigenvalue is achieved when 
$\sigma^{x}_{im1}\otimes\sigma^{x}_{im2}\otimes\sigma^{x}_{im3}$ is replaced by 1 (i.e. the 
ancilla qubits are in $(\ket{000}+\ket{111})/\sqrt{2}$. $H_{anc}$ has restricted the ancilla 
qubits to be in the subspace spanned by $\ket{000}$ and $\ket{111}$), and we effectively 
recover $V(s_i)$. We can see that the purpose of those $\sigma_{im1}\ldots$ terms is to 
enforce the product relation among the $\{B_{im1},B_{im2},B_{im3}\}$. Notice the first excited 
state is also the same as in $V(s_i)$ because the B terms are positive definite, so if the 
ancilla qubits are in $(\ket{000}-\ket{111})/\sqrt{2}$, the increase in energy would be more 
than that of the excited states due to the computational qubits.

Under what condition will this procedure be inefficient? Note that in (\ref{vprime}), the 
original $B_{im1}$ terms are multiplied by $\delta^{-2}$. The local reduction requires the 
approximating terms to be very large compared to other terms in the Hamiltonian. If reduction 
is later applied repeatedly to terms coupling to ancilla qubits from the previous steps, the 
energy level required for the reduction scales exponentially. When we normalize the total 
energy to unity, this equivalently means the gap between the ground state and the first 
excited state shrinks exponentially. Such repeated approximation could be useful when we need 
to implement a shallow circuit with the noise resistant properties of the adiabatic 
computation model and the restriction of 2-local interaction. For example, in conjunction with 
teleportation circuits, the repeated approximation may not be necessary as we can teleport 
many independently and adiabatically prepared unitary operations. For a generally efficient 
mapping, we would need either a procedure that directly reduces an n-local Hamiltonian 
to a 2-local approximation, subjected to the constraints described in section 4, or some kind 
of adaptive mapping that exploit structures of specific circuits. 

In the next section, we will see how similar repeated use of the three-qubit gadget can give 
rise to an efficient adiabatic algorithm once we relax the model.

\section{6. Generalized adiabatic algorithms and Holonomic Quantum Computing}

As mentioned in the introduction, if we are willing to relax the property that the Hamiltonian 
keeps track exactly what the correct state is, we will have more freedom to design the 
Hamiltonian. Going back to the construction in section 1, it is clear that we need different 
Hamiltonians for the \em same \em quantum gate at different stages of the computation. Yet we 
know that the unitary transformation due to the application of a time-dependent Hamiltonian 
over a period of time, $U = \cal{T}$$exp(i\int_0^T H(t)dt)$ is independent of the state, so 
why do we need different Hamiltonians for different stages? The reason is that we have so far 
ignored the phases of the transformation due to the adiabatic evolution. The phases include 
both dynamical and geometric components:
\be
\phi_n(T)=exp(-i\int_0^T E_n(t)dt+i\gamma_n(T))
\ee
where
\be
\gamma_n(T)=i\int^T_0 \bra{n,t}{d \over dt}\ket{n,t} dt
\ee
is the geometric phase \footnote{This is not a closed path and thus is not gauge invariant, as 
pointed out in \cite{openpath2}. We will nonetheless keep the terminology. To readers 
unfamiliar with Berry's phase, the word "gauge" here refers to the U(1) degree of freedom 
associated with a normalized eigenvector in the Hilbert space} and $E_n(t), \ket{n,t}$ are the 
eigenvalues and eigenvectors of $H(t)$. Therefore, if we naively apply $U(t) H_0 U^\dag(t)$ to 
a state $\p$ without making sure that $\p$ is an eigenstate of $H_0$, relative phases can 
develop between the eigenstates superposed to form $\p$. It is not difficult to cancel out the 
relative dynamical phases - all we need to do is to apply $-H_0$ for the same period of time 
(or modulo $2\pi$). For the geometric phases, they are often ignored in an open path evolution 
since they can be gauged away by choosing a different set of basis. However, the moment we 
decompose $\p$ into eigenstates of $H(0)$, the gauge is fixed; if we choose a different set of 
basis at some other time, we would not obtain $U\p$ at the end. Therefore the open-path 
geometric phase must be taken into account and cancelled accordingly. This gives rise to the 
following partially adiabatic algorithm:

To apply two-qubit gate U from a circuit: \\
1. Pick a simple, 1-local Hamiltonian $H_0$, $\| H_0 \| \leq$ 1. \\
2. Apply the ${1\over 2}(I+U(t)H_0U(t)^\dag)$ at any rate from t=0 to T; $U(0)=I$ and $U(T)=U$ 
\\
3. Calculate $G$ such that $G\ket{n}=\gamma_n\ket{n}$. \\
4. Apply ${1\over 2}(I-H_0+2G/T)$ for time T. \\

This algorithm, of course, does not enjoy property II) for all time because of step 4. 
But if we make good use of the transformation due to the geometric phase, such that, for 
example, $e^{iG}U(T)=U$ (which is nontrivial to solve since G depends on the path $U(t)$), the 
G term can be dropped from step 4, and the cancellation can be greatly simplified - in fact 
the cancellation would be the same whatever gate we want to implement.

This is also reminisicent of the Geometric Quantum Computation model \cite{geom}, which uses 
the Abelian geometric phase to implement each gate and requires the cancellation of dynamical 
phases. The difference is that our algorithm  uses an open path and thus involves 
non-geometric components. 

Next let us consider how we can avoid having to cancel the phases. This is only possible if 
any state we want to apply the quantum gate on is an eigenstate of $H_0$. But section 4 tells 
us this state cannot be a unique eigenstate without $H_0$ becoming n-local in general, so we 
will have to deal with degenerate states. Without any knowledge about the state, we would have 
to make all $2^n$ n-qubit states degenerate - but this means $H_0$ acts trivially on all 
qubits! The dilemma is solved by adding ancilla qubits - we can arrive at a non-trivial 
$U(t)H_0U(t)^\dag$ if $U(t)$ couples between the computational qubits and the ancilla qubits. 
Notice that $U(0)H_oU(0)^\dag = U(T)H_0U(T)$ since both are trivial on computational qubits, 
so the Hamiltonian goes through a cycle. All the relative phases are accumulated between 
states corresponding to the ancilla qubits and do not affect our calculation. We have thus 
arrived at the Holonomic Quantum Computation (HQC) model\cite{holo1}. 

In order to have a non-trivial control manifold, we can use qutrits with states $\ket{0}$ and 
$\ket{1}$ as the usual qubit states and $\ket{2}$ for control, or we can add ancilla qubits. 
In general, identifying the manifold that has the right holonomy group and finding the path 
for each two-qubit transformation in the circuit model is very difficult. Recently, Tanimura 
et al \cite{holo2} settled the mathematical question of finding a shortest path given an 
arbitrary holonomy group element in a homogenous bundle, which implies that with the addition 
of just one ancilla qubit, we can implement any two-qubit transformation in the space of 
computational states. Let us consider the implementation of a CNOT gate as an example of this 
result. Take the Hamiltonian on the ancilla qubit to be 
\[
H_0 = \left(\begin{array}{cccc}
E_0 &0  \\
0 & E_1 \\
 \end{array}\right)	
\] 

We can write the time-dependent Hamiltonian, including the two qubits to be transformed, as
\be
H(t)=E_1 e^{tX}V_1 V_1^\dag e^{-tX} + E_0 e^{tX}V_0 V_0^\dag e^{-tX}; t\in[0,1]
\ee
where $V_0$, $V_1$ are 8x4 matrices:
\[
V_0 = \left(\begin{array}{cccc}
I_4  \\
 0 &  \\
 \end{array}\right)	
V_1 = \left(\begin{array}{cccc}
0  \\
I_4 &  \\
 \end{array}\right)	
\] 
$I_4$ denotes the 4x4 identity matrix. We start by preparing the ancilla qubit in ground state 
$E_0$. To implement a CNOT optimally, \cite{holo2} found $X$ to be:
\[
X = \left(\begin{array}{cccc}
A & B \\
-B^\dag & 0 \\
 \end{array}\right)	
\] 
where A and B are:
\[
A = i \pi \left(\begin{array}{cccc}
2&0&0&0 \\
0&2&0&0\\
0&0&1&1 \\
0&0&1&1\\
 \end{array}\right)	
B = {i \pi \over \sqrt{2}} \left(\begin{array}{cccc}
0&0&0&0 \\
0&0&0&0 \\
0&0&0&-1 \\
0&0&0&1 \\
 \end{array}\right)	
\]

We can see that the Hamiltonian only acts on the two computational qubits and the one ancilla 
qubit, so other computational qubits are not affected at all. To do the same computation using 
2-local Hamiltonians, we can now apply the three-qubit gadgets of \cite{kkr} described 
earlier. At the end of each cycle, unlike the case in section 5, the ancilla qubits for the 
reduction can be discarded and reused in the next step. The total number of ancilla qubits 
required is three times the number of terms in $\sum_m B_{im1}B_{im2}B_{im3}$ of 
(\ref{vform2}), which is a constant. Following the same analysis, if we want the final state 
to be accurate up to $O(\epsilon)$, the allowed error at each step should be $O(\epsilon/L^2)$ 
where $L$ is the depth of the circuit. The energy gap required is thus $O(\epsilon^3/L^6)$ and 
the running time is $O(L^{12}/\epsilon^6)$. Note that this bound may be far from tight, and it 
is quite possible that most circuits can be implemented in far shorter time. In any case, we 
have arrived at a fully adiabatic evolution that computes efficiently any problem solvable by 
quantum circuit using only 2-local Hamiltonians and a constant number of ancilla qubits.

\section{7. Connection to the history approach}

Comparing the above running time estimate to the results in \cite{adkll,kkr}, which we will 
refer to this as the history approach, it appears that the latter still seems superior even 
though it produces the entire history of the computation instead of only the final output 
(This just means one has to repeat the process $O(L)$ times to get the final output or put in 
identity gates.) It is natural to ask what is special about this approach and how it is 
related to the models we have studied. For a circuit of n qubits and L gates, the history 
approach has the following final state:
\be
\ket{\psi_{final}}={1\over \sqrt{L+1}}\sum_{l=0}^L 
U_lU_{l-1}...\ket{0}\otimes\ket{1^l0^{L-l}}^c
\label{history}
\ee  
where $\ket{1^l0^{L-l}}^c$ denotes the state of L ancilla qubits serving as a clock. We refer 
the readers to \cite{adkll} for the general form of the 3-local Hamiltonian which has the 
above state as its unique ground state. Let us just look at the simplest example - a circuit 
with one two-qubit gate U and n computational qubits. The initial and final states are:
\begin{eqnarray} 
\ket{\psi_{ini}}=\ket{0}\otimes\ket{0}^c; \\ 
\ket{\psi_{final}}={1 \over \sqrt{2}}(\ket{0}\otimes\ket{0}^c+U\ket{0}\otimes\ket{1}^c) 
\end{eqnarray}
where one ancilla qubit suffices for the clock. The corresponding Hamiltonians are:
\begin{eqnarray}
H_{ini}=\ket{1}\bra{1}^c \\
H_{out}={1\over2}(I\otimes\ket{0}\bra{0}^c+I\otimes\ket{1}\bra{1}^c \nonumber \\
 - U\otimes\ket{1}\bra{0}^c - U^\dag\otimes\ket{0}\bra{1}^c)
\end{eqnarray}
Actually we have deliberately omitted a term $\sum_{n}\ket{1}\bra{1}\otimes\ket{0}\bra{0}^c$ 
that would make $\ket{\psi_{ini}}$ the unique ground state of $H_{ini}$. This does not change 
our analysis as long as we prepare the state in $\ket{\psi_{ini}}$ at the beginning. In 
\cite{adkll} this is the $H_{input}$ term, which does not affect the evolution equation 
because it remains zero for all time. Aharonov et al shows - for this case, trivially - that 
slowly interpolating from $H_{ini}$ to $H_{final}$ takes $\ket{\psi_{ini}}$ to 
$\ket{\psi_{final}}$. Notice that the Hamiltonians are highly degenerate in the sense that if 
we replace $\ket{\psi_{ini}}$ and  $\ket{\psi_{final}}$ by
\begin{eqnarray}
\ket{\psi_{ini}^{'}}=V\ket{0}\otimes\ket{0}^c; \\ 
\ket{\psi_{final}^{'}}={1 \over \sqrt{2}}(V\ket{0}\otimes\ket{0}^c+UV\ket{0}\otimes\ket{1}^c) 
\end{eqnarray}
for some U(n) unitary operator V, the same analysis would go through. This gauge freedom is 
already reminiscent of HQC; let us make the connection more explicit. Define the 
time-dependent Hamiltonian 
\begin{eqnarray}
H(t)={1\over2}(P(t)I\otimes\ket{0}\bra{0}^c+Q(t)I\otimes\ket{1}\bra{1}^c \nonumber \\
 - R(t)U\otimes\ket{1}\bra{0}^c - R(t)U^\dag\otimes\ket{0}\bra{1}^c)
\end{eqnarray}
such that P=R=0, Q=2 gives $H_{ini}$ and P=Q=R=1 gives $H_{final}$. It is easy to see that
\be
\ket{\psi(t)}=\{normalization\}[\alpha(t)\ket{0}\otimes\ket{0}^c+\beta(t)U\ket{0}\otimes\ket{1
}^c]
\ee
is a ground state as long as $PQ=R^2$ and $P\alpha=R\beta$. Suppose we have completed the 
original evolution and obtained $\ket{\psi_{final}}$ ($\alpha, \beta=1$), but decided to keep 
going. We can slowly change from P=Q=R=1 to P=2,Q=R=0 while maintaining $PQ=R^2$ (this helps 
keeping a large gap), and we would arrive at $U\ket{0}\otimes\ket{1}^c$ ($\alpha=0,\beta=1$). 
This is essentially a holonomic cycle if we adiabatically rotate the clock qubit back to 
$\ket{0}^c$. In practice we can just relabel the clock qubit if we want to do this repeatedly.

This gives rise to the interpretation that the history approach is in fact half of a holonomic 
cycle. The cycle we have just shown is very similar to the CNOT example in section 6, as it 
requires a 3-local Hamiltonian, which too can be reduced by the three-qubit gadget, and one 
ancilla qubit; it is merely a different path in the control manifold. Repeated application of 
the gadget allows efficient universal quantum computing, as described in section 6. 

We can take this further and interpret the history approach for L quantum gates as part of a 
holonomic cycle with a control manifold augmented with the space of L ancilla qubits. For the 
3-local Hamiltonian case, this is not difficult to see. Suppose we have starting Hamiltonian 
$H_i$ and final Hamiltonian $H_f$, such that linear adiabatic evolution takes starting state 
\be
\psi_i = \ket{0}\otimes\ket{0}^c
\ee
to
\be
\psi_f = {1\over\sqrt{L+1}}\sum_{i=0..L} U_i...U_0 \ket{0}\otimes\ket{i}^c
\ee
where $U_0=1$ and we have simplified the notation for the clock. These Hamiltonians can be 
constructed according to \cite{adkll}, except we omit the term $H_{input} = \sum_i 
\ket{1}\bra{1}_i \otimes \ket{0}\bra{0}^c$ so that we preserve the gauge freedom mentioned 
above. This does not affect the adiabatic evolution because the Hamiltonian does not couple 
between the degenerate spaces. Now we can similarly construct another pair of Hamiltonian 
$H_i'$ and $H_f'$ corresponding to a reverse circuit $U_L^{\dag},U_{L-1}^{\dag}...$ such that 
it takes 
\be
\psi_i' = U_L...U_0 \ket{0}\otimes\ket{0}^c
\ee
to
\be
\psi_f' = {1\over\sqrt{L+1}}\sum_{i=0..L} U_{L-i}...U_0 \ket{0}\otimes\ket{i}^c
\ee
Simple inspection shows that $\psi_f$ and $\psi_f'$ are the same up to relabelling of clock 
qubits. Therefore, if we implement the evolution $H_i \to H_f(H_f') \to H_i'$ up to relabelled 
clocks, we can take
\be 
\psi_i = \ket{0}\otimes\ket{0}^c
\ee
to
\be
\psi_i' = U_L...U_0 \ket{0}\otimes\ket{0}^c
\ee
and thus complete a holonomic implementation of the L gates. This opens up the possibility of 
further optimizing the path; a more complete analysis is left for the future.

\section{Discussion}

We have looked at various forms of adiabatic quantum computatation and studied their resource 
requirements as well as possible approximations. One issue we have not addressed at all is 
what noise resistant properties different models can have. It would appear that the direct, 
non-holonomic approach in section 2 places stronger condition on the states, as any deviation 
requires higher energy, whereas the degenerate states in the holonomic approach have no 
protection against transition within the degenerate level. This generalization may however be 
simplistic; a better analysis should be with respect to particular experimental 
implementations. We simply hope that this paper has provided a more unifying picture of 
adiabatic algorithms that will eventually lead to a toolbox experimentalists can refer to for 
different specific applications. 

There are at least a few directions for further studies. 1) In section 2 we constructed an 
adiabatic equivalent of any arbitrary circuit. What remain unclear are the general properties 
of such Hamiltonians, namely those of the form $UH_0U^\dag$ where $H_0$ is a simple 
Hamiltonian and U is a polynomial-sized circuit. If it is n-local, in what sense is it simpler 
than the most general Hamiltonian? Are there things about a circuit that we can learn through 
this corresponding Hamiltonian? 2) We discussed in section 4 various constraints on 
approximating $UH_0U^\dag$ with k-local terms. The general picture is still unclear; it would 
be useful to understand precisely under what condition an n-local Hamiltonian can be 
approximated by a 2-local one. 3) We have proposed one partially adiabatic algorithm in 
section 6 and we have suggested that the adiabatic construction in section 5 could be useful 
as a small section of a larger algorithm. It is interesting to investigate what merits, if 
any, these partially adiabatic algorithms possess. 4) For HQC, there should be room for 
improvement. If we implement one two-qubit gate at a time, it is important to see how the use 
of three-qubit gadget can be optimized and how a tighter bound on the evolution time can be 
obtained. At the same time, the history approach discussed in section 7 may lead to new class 
of HQC methods that implement many gates together efficiently. 5) In order to build an 
efficient computation model using only 2-local Hamiltonian and adiabatic evolution, we have 
been naturally led to the use of non-Abelian geometric phase. The use of holonomy, however, 
may not be the only option. For example, open path non-Abelian geometric phase \cite{openpath} 
can be non-trivial as well and may lead to novel ways of implementing multiple qubit gates. 
These questions are beyond the scope of this paper, and we believe that adiabatic quantum 
computing remains an exciting area to explore.

\section{Acknowledgement}

M.S. would like to thank Julia Kempe, Geordie Rose, Yaoyun Shi and Colin Williams for valuable 
discussions and crucial correspondences at various stages of the preparation of this work.


\end{document}